\documentclass[aps,english,prl,amsmath,amsfonts,amssymb,superscriptaddress,twocolumn,citeautoscript]{revtex4}
\usepackage[T1]{fontenc}
\usepackage[latin9]{inputenc}
\usepackage{amsmath}
\usepackage{amssymb}
\usepackage{wasysym}
\usepackage{esint}
\usepackage{graphicx}
\usepackage{times}
\usepackage{nicefrac}
\usepackage{appendix}
\usepackage{textcase}
\usepackage{subfigure}
\usepackage{empheq}
\usepackage{color}
\usepackage{leftidx}
\usepackage{makecell}
\usepackage{stackrel}
\makeatletter
\@ifundefined{textcolor}{}
{%
 \definecolor{BLACK}{gray}{0}
 \definecolor{WHITE}{gray}{1}
 \definecolor{RED}{rgb}{1,0,0}
 \definecolor{GREEN}{rgb}{0,1,0}
 \definecolor{BLUE}{rgb}{0,0,1}
 \definecolor{CYAN}{cmyk}{1,0,0,0}
 \definecolor{MAGENTA}{cmyk}{0,1,0,0}
 \definecolor{YELLOW}{cmyk}{0,0,1,0}
}

\renewcommand{\eqref}[1]{Eq.~\ref{eq:#1}}

\newcommand{\figref}[1]{Fig.~\ref{fig:#1}}

\makeatother

\usepackage{babel}
\begin{document}
\title{Inverse design of lightweight broadband  reflector for efficient lightsail propulsion}
\author{Weiliang Jin}
\affiliation{Department of Electrical Engineering, Ginzton Laboratory, Stanford University, Stanford, California 94305, USA}
\author{Wei Li}
\affiliation{Department of Electrical Engineering, Ginzton Laboratory, Stanford University, Stanford, California 94305, USA}
\author{Meir Orenstein}
\affiliation{Department of Electrical Engineering, Technion-Israel Institute of Technology, 32000 Haifa, Israel}
\author{Shanhui Fan}
\affiliation{Department of Electrical Engineering, Ginzton Laboratory, Stanford University, Stanford, California 94305, USA}

\begin{abstract}
  Light can exert forces on objects, promising to propel a meter-scale
  lightsail to near the speed of light. The key to address many
  challenges in such an ambition hinges on the nanostructuring of
  lightsails to tailor their optical scattering properties. In this
  letter, we present a first exhaustive study of photonic design of
  lightsails by applying large-scale optimization techniques to a
  generic geometry based on stacked photonic crystal layers. The
  optimization is performed by rigorous coupled-wave analysis amended
  with automatic differentiation methods for adjoint-variable gradient
  evaluations. Employing these methods the propulsion efficiency of a
  lightsail that involves a tradeoff between high broadband
  reflectivity and mass reduction is optimized. Surprisingly,
  regardless of the material choice, the optimal structures turn out
  to be simply one-dimensional subwavelength gratings, exhibiting
  nearly 50\% improvement in acceleration distance performance
  compared to prior studies. Our framework can be extended to address
  other lightsail challenges such as thermal management and propulsion
  stability, and applications in integrated photonics such as compact
  mirrors.
\end{abstract}
\maketitle
Light can exchange momentum with objects~\cite{novotny2012principles},
leading to many vital breakthroughs in the field of nanotechnology
such as optical tweezers for precise manipulation of nanoscale
particles~\cite{moffitt2008recent,gao2017optical}.  Optical force can
also play a crucial role in much larger lengthscale applications such
as space travel, including the recent launch of solar sail driven by
sun light~\cite{mansell2020orbit}. Another and even more ambitious
project is the Starshot Breakthrough Initiative that aims to
accelerate a meter-size spacecraft to 20$\%$ of the speed of light, so
that it can reach a nearby galaxy Proxima Centauri in 20
years~\cite{lubin2016roadmap,parkin2018breakthrough}. By far, the most
plausible propulsion mechanism is based on optical
force~\cite{kulkarni2018relativistic}, or radiation pressure from
GW/m$^2$ level lasers ~\cite{atwater2018materials}.

While such a project requires multidisciplinary
efforts~\cite{parkin2018breakthrough} such as materials
science~\cite{moura2018centimeter}, mechanical engineering,
astrophysics~\cite{lingam2020propulsion}, and
telecommunications~\cite{bird2020advances}, many key challenges can be
alleviated via probing the boundary of photonic design, including
efficient propulsion~\cite{atwater2018materials}, heat
management~\cite{ilic2018nanophotonic}, laser beam
focusing~\cite{noyes2019analyzing}, and self-stabilization
~\cite{ilic2019self,siegel2019self,manchester2017stability,myilswamy2020photonic}.
They all contain many tradeoffs that were so far optimized by tuning
few geometric parameters of simple photonic structures, leaving
possibly much room for improvement by systematically studying more
complicated structures. An important tool for accomplishing this task
is inverse design method capable of exploring millions of design
variables that have been introduced into photonics in the last
decade~\cite{molesky2018inverse,yao2019intelligent}, proving to be
powerful in discovering structures whose performances hit theoretical
bounds~\cite{angeris2019computational}, or in suggesting the existence
of tighter fundamental limits, e.g. recently improved bounds on
optical absorption and scattering cross
sections~\cite{molesky2020t}, optical
force~\cite{lee2017computational},
near-field~\cite{venkataram2020fundamental,jin2019material} and
far-field thermal radiation~\cite{molesky2019t}.


In this letter, as an initial step towards systemically pushing
forwards photonics-related performances of lightsails, we apply
large-scale optimization methods to identify lightsail geometric
design criteria for optimal propulsion efficiency, crucial to lowering
both laser power and phase array
size~\cite{ilic2018nanophotonic}. More specifically, we seek to
minimize a figure-of-merit (FOM) described by \eqref{D}, known as
acceleration distance that involves a tradeoff between broadband
reflectivity and lightsail mass. Previous optimizations were based on
simple photonic crystal (PhC)
slabs~\cite{atwater2018materials,ilic2018nanophotonic}, in contrast,
here we explore a generic class of geometries, stacked PhC layers
whose dielectric spatial profiles can be arbitrarily set within the
unit cell.
Gradient-based optimization methods are applied to simultaneously
optimize over dielectric distributions, periodicity of PhC, and
thickness of each layer, whose gradients are conveniently evaluated
with automatic differentiation
methods~\cite{minkov2020inverse}. Different constituent materials of
lightsails and payload mass are studied. Surprisingly, we demonstrate
that for both high-index material such as silicon, and lower-index
material such as silicon nitride, the optimal structure converges to a
one-dimensional (1D) subwavelength grating, a robust solution against
a wide range of payload mass. The FOM of this optimal structure
exhibits a nearly $50\%$ improvement against that of previously
explored structures. The enhancement is attributed to the destructive
interference of two guided modes that can be supported in such a
grating of small material volume filling ratio and at a subwavelength
thickness. The optimal solutions can converge to more complicated
two-dimensional (2D) PhC structures when more emphasis is imposed on
maximizing reflectivity, e.g. when the payload mass is large. Finally,
we conclude that in general with optimizations, high-index material
yields better performance.

In a general context, a broadband mirror is traditionally realized
with a metallic reflector, which nevertheless possesses high material
absorption loss, incompatible with the thermal management requirement
of the Starshot projects~\cite{atwater2018materials}. Recent advances
in nanotechnologies have led to the development of low-loss
all-dielectric mirrors such as the distributed Bragg
reflector~\cite{coldren1997diode}, and several more compact schemes.
One representative class of structures is derived from metamaterial
principles by exploiting the single-negative response, commonly
implemented with a single layer of microspheres, cubes
~\cite{slovick2013perfect,moitra2014experimental,moitra2015large}, or
shapes dicovered with machine learning
methods~\cite{harper2020inverse}. Another direction is based on
guided-wave analysis that employs the double-mode destructive
interference effect, realized with a subwavelength 1D grating of
high-index contrast
dielectrics~\cite{mateus2004ultrabroadband,karagodsky2010theoretical}. The
mechanism of the two approaches can be unified~\cite{ko2018wideband}.
However, a comprehensive study on the mirror design principle for
minimal mass is missing, a principle relevant in many integrated
photonic applications, and in efficient propulsion of lightsails.

\begin{figure}[htbp]
  \centering
  \includegraphics[width=0.7\columnwidth]{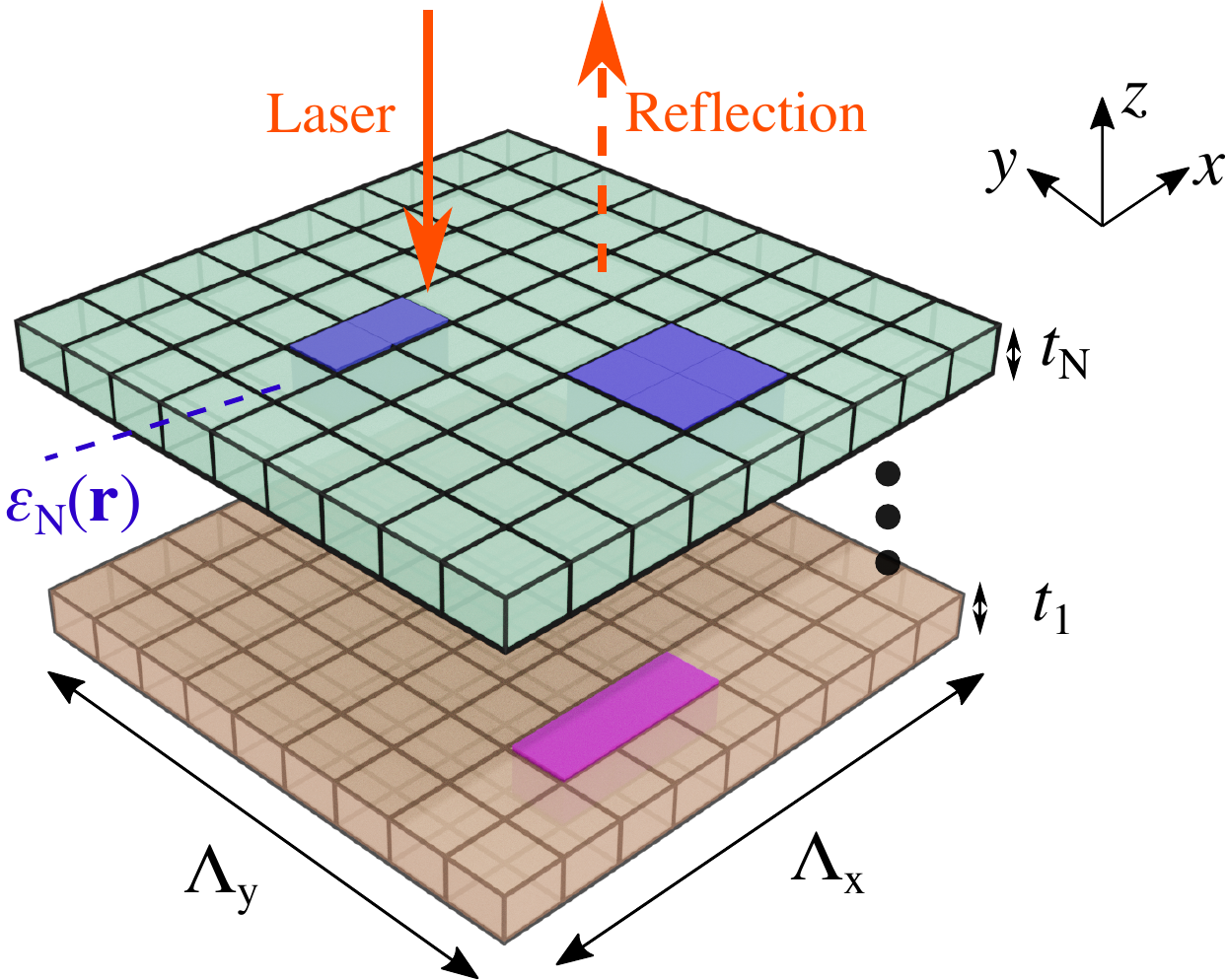}
  \caption{Schematic of a lightsail propelled by laser beams. The
    lightsail consists of stacked photonic crystal layers of period
    $\Lambda_{x(y)}$. The design space includes the period
    $\Lambda_{x(y)}$, the in-plane dielectric index value
    $\varepsilon_i(\mathbf{r})$ at each grid point, and the thickness
    $t_i$ of the i-th layer. Here different colors represent different
    materials.}
  \label{fig:scheme}
\end{figure}

As shown in \figref{scheme}, we envision a spacecraft consisting of a
payload (not shown) and a lightsail, where the latter is structured to
enhance the optical force along the normal ($-z$) direction exerted on
the sail by an incident high-power laser beam. From momentum
conservation, the optical force increases with lightsail's
reflectivity ~\cite{chen2011optical,ilic2018nanophotonic}. A practical
FOM characterizing the propulsion efficiency is the distance $D$ for
the spacecraft to be accelerated to a target velocity, which takes into
account the trade-off between optical force and kinetic quantities
such as mass, which is captured by the following
equation~\cite{atwater2018materials,kulkarni2018relativistic},
\begin{equation}
D = \frac{c^3}{2I}(\rho_l+\rho_s)\int_0^{\beta_f}\mathrm{d}\beta  \frac{h(\beta)}{R[\lambda(\beta)]} \label{eq:D}
\end{equation}
where $\rho_{s(l)}$ is the area density mass of the lightsail
(payload), $I$ the laser intensity, $c$ the speed of light, and
$h(\beta)=\beta/(1-\beta)^2\sqrt{1-\beta^2}$ encodes relativistic
factors depending on velocity fraction $\beta=v/c$. The integration is
performed from stationary motion to a target velocity $c\beta_f$, and
during this time interval the laser wavelength $\lambda$ in the
lightsail frame is Doppler redshifted from $\lambda_0$ to
$\lambda(\beta)=\lambda_0\sqrt{(1+\beta)/(1-\beta)}$. A representative
value as in the Starshot project is
$\lambda(\beta_f)\approx1.22\lambda_0$, revealing that the
reflectivity $R$ of the lightsail needs to be enhanced over a large
bandwidth. Minimization of $D$ has a direct impact on reducing both
the size of the laser phase array that needs to account for
diffraction~\cite{ilic2018nanophotonic}, as well as the total power
consumption.

\emph{Optimization.} To minimize $D$, we seek to structure the
lightsail with wavelength or subwavelength features to tailor its
optical scattering properties. Since the lightsail should involve two
vastly different geometric scales, a macroscopic area $\sim 10$~m$^2$
and nanoscale thickness on the order of $ 10^2$~nm, the suitable
generic class of geometries is stacked PhC slabs. As shown in
\figref{scheme}, each layer $i$ with thickness $t_i$ is uniform along
$z$-direction, and periodically structured in the $xy$-plane. Such a
platform contains a rich library of geometries, including the
aforementioned dielectric broadband mirrors, and previously explored
lightsail structures such as uniform slabs, multilayer stacks, and PhC
pillars or
holes~\cite{atwater2018materials,ilic2018nanophotonic}. Inverse design
of more complicated geometries such as aperiodic structures including
Moir\'e lattices and photonic quasicrystals for inherent broadband
responses~\cite{dal2013optics}, and curved surfaces for mechanical
stability~\cite{manchester2017stability}, will be considered in a
future work.

In order to probe the limit of $D$ via photonic designs, we apply the
``topology'' optimization approach~\cite{molesky2018inverse} that
enables us to explore the largest possible design space. More
specifically, we discretize the unit cell of the $i$-th layer into
$M_i\times M_i$ grids, and allow to choose between materials at each
grid point independently. This contributes to at least
$\sum_{i=1}^NM_i^2$ design variables, each of which can take $n$
discrete values, where $N$ is the total number of layers, and $n$ the
number of candidate materials. We also treat the periodicity and the
thickness of each layer as additional independent variables. The key
to the tractability of such large-scale optimizations is the use of
gradient-based optimization algorithms, such as the method of moving
asymptotes~\cite{svanberg2002class}. To make use of these approaches,
the index of refraction at each grid can initially vary continuously
between various types of materials, and subsequently be binarized with
filter and regularization methods~\cite{jensen2011topology}. However,
such local optimization algorithms are known to be ill-behaved over
high-index dielectric structures~\cite{liang2013formulation}, due to
the presence of many narrow-band resonances such as bound states in
the continuum~\cite{hsu2016bound}. To better approach globally optimal
solutions, we employ a relaxation method that broadens any high-Q
response by adding fictitious material absorption loss to the entire
system
, and eventually turning it off~\cite{liang2013formulation}. By
employing this approach, our optimization results are highly
insensitive to initial parameters, a hint for possibly globally
optimal results.


The primary complexity of inverse design problems lies in the
derivation of the adjoint-variable problem for efficient gradient
evaluation~\cite{molesky2018inverse}. While our FOM in \eqref{D} is
simple, future work in optimizing other aspects of lightsails may lead
to convoluted FOMs that makes the derivation of gradient information
mind-twisting. Thankfully, recent advancement in machine learning
community has led to the development of various convenient packages
for automatic differentiation, the application of adjoint-variable
methods to arbitrary computational
graphs~\cite{minkov2020inverse}. With those tools, we only need to
implement forward problems, while the backward gradient evalulations
will be generated automatically.
For efficient optimization of \eqref{D} over arbitrarily structured
PhC layers, we have implemented a package~\cite{grcwa} that extends
rigorous coupled-wave analysis (RCWA) ~\cite{liu2012s4} with the
automatic differentiation software
Autograd~\cite{maclaurin2015autograd}.
\begin{figure*}[htbp]
  \centering
  \includegraphics[width=2\columnwidth]{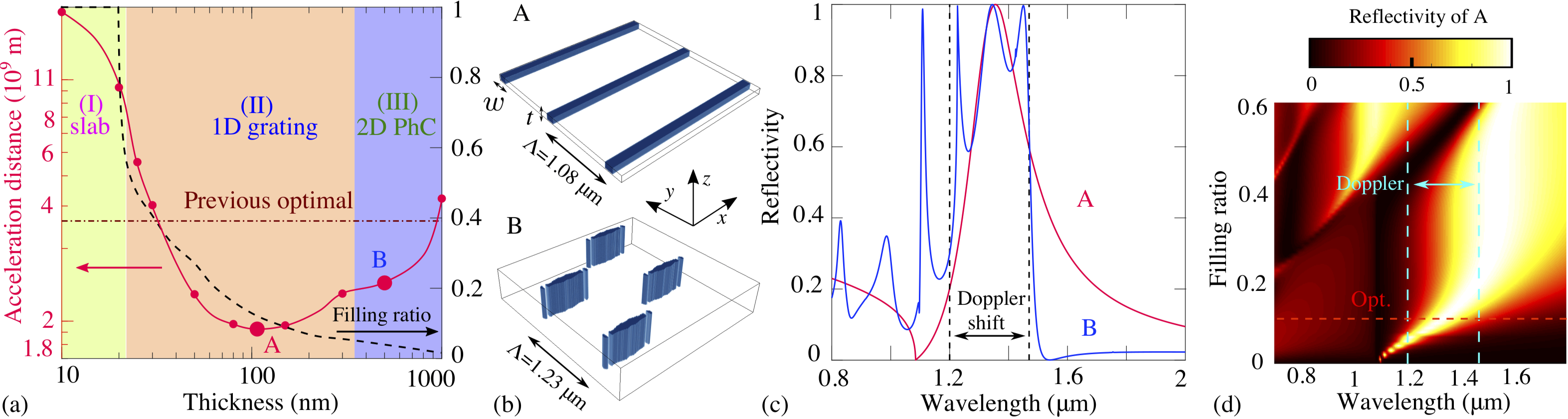}
  \caption{Inverse design of lightsail made of crystalline
    silicon. (a) acceleration distance (left axis) and filling ratio
    (right axis) of the structures optimized at each thickness,
    illustrate the transition of optimal geometry from uniform slabs
    (green), one-dimensional gratings (orange), to two-dimensional
    PhCs (purple). Red dash-dotted line denotes the previous optimal
    design~\cite{atwater2018materials}. Two representative structures
    at points $A$ and $B$ are illustrated in (b), and their calculated
    reflection spectrum is shown in (c). The $x$-polarized laser beam
    is propagating along $-z$ direction. (d) reflectivity of the
    optimal grating structure $A$ as a function of wavelength and
    width $w$, or the filling ratio $w/\Lambda$. Typical Starshot
    parameters~\cite{atwater2018materials} are assumed: payload mass
    $0.1$~g, lightsail area $10$~m$^2$, and laser intensity
    $10$~GW/m$^2$.}
  \label{fig:silicon}
\end{figure*}

\emph{Results.} We apply the optimization formulation to identify the
optimal structuring criteria that minimize $D$. A glimpse on \eqref{D}
indicates that design criteria depend on the refractive indices of the
constituent materials that dictate scattering properties, and payload
mass that governs the degree of tradeoff between reflectivity and
lightsail mass. To address those possibilities, we explore two types
of dielectric materials of distinct refractive indices, as well as
various values of payload mass. Without further specification, we
assume typical Starshot parameters in which a laterally uniform,
linearly polarized laser beam of intensity $10$~GW/m$^2$ and
wavelength $\lambda_0=1.2$~$\mu$m is incident normally on a lightsail
of area $10$~m$^2$. To allow large design space, we consider fine grid
size $\lesssim 10$~nm, leading to at least $10^{4}$ spatial design
variables per $1$~$\mu m^2$ in each layer. We consider two choices of
material on each grid point: the target material and a vacant space
with unity refractive index and negligible mass, which in practice can
be vacuum or aerogels~\cite{jo1997application}. In case of vacuum, a
low-index substrate is needed for mechanical rigidity, which can be
treated as payload mass.

We begin by studying the optimizations of a representative high-index
material, crystalline silicon. To gain insights into optimization
results, we start by treating the overall thickness as a
hyperparameter, namely, optimizing over period and material
distributions at each thickness independently. The FOM (red solid
line) and material volume filling ratio (black dashed line) of the
optimal structures are summarized in \figref{silicon}(a), uncovering
three distinct regimes of structural choices. First, at small
thickness $\lesssim 20$~nm, as may be expected, the optimization
converges to a finite-thickness uniform slab (green region) since at
deep subwavelength thicknesses, the reflectivity of high-index
material increases more steeply than mass with filling ratio.

Second, for the intermediate thickness range we obtain the globally
optimal solution. Surprisingly, even though we are optimizing over
multiple independent layers of $N\gtrsim 2$ and material distributions
on 2D grids, the optimal shape is a 1D grating (orange region),
depicted in the upper \figref{silicon}{b}. With increasing thickness
that allows for enhanced scattering, the optimal filling ratio,
$w/\Lambda$, decreases in favor of lighter mass, with simultaneously
enhanced average reflectivity in the Doppler-shift bandwidth,
resulting in dramatically decreasing $D$ that exhibits a minimum at
thickness $107$~nm (denoted as Grating $A$ in
\figref{silicon}(a)). Furthermore, we observe that while the optimal
period $\Lambda$ also varies with thickness, it is always
subwavelength $< \lambda_0$, e.g.  $\Lambda=1.08$~$\mu$m for Grating
$A$. Such a subwavelength grating eliminates the diffraction in
off-normal directions, which is important for enhancing the optical
force along the normal direction. Compared to previously explored
structures~\cite{atwater2018materials}, the acceleration distance of
Grating $A$ is $D_{\mathrm{Si}}\approx 1.9\times 10^9$m, which
represents a nearly $50\%$ improvement.

In another context, 1D gratings have been proposed as broadband
reflectors, known as high-contrast subwavelength
gratings~\cite{mateus2004ultrabroadband}. They have been demonstrated
to achieve nearly $100\%$ reflection over a bandwidth
$\Delta\lambda/\lambda\gtrsim 30\%$ for either TE or TM polarized
light, arising from the destructive interference of two guided modes
that prohibits transmission. We examine if similar mechanism is the
source of the performance of our optimized Grating $A$ by plotting its
reflection spectrum in \figref{silicon}(c). A Fano resonance feature
is visible near the Doppler shift range, demonstrating that the high
broadband reflectivity is indeed attributed to the double-mode
interference effects. Another important observation is that for the
incident light polarized along the $x$-axis, denoted in
\figref{silicon}(b), the algorithm always finds a grating extending
along the same $x$-direction. This is consistent with previous
studies~\cite{karagodsky2010theoretical} that gratings parallel to the
light polarization direction, when compared to those of perpendicular
orientations, can achieve broadband reflection with smaller filling
ratio, and consequently lighter mass. To gain more insights into the
optimization process, in \figref{silicon}(d), we show the reflectivity
plot of Grating $A$ as a function of wavelength and filling ratio for
a fixed thickness and period. Indeed, high reflection mostly occurs in
the non-diffractive region, $\lambda>\Lambda$. As the filling ratio
decreases, the bandwidth of the high reflection region shrinks, a
typical tradeoff between the mass and average reflectivity. To
minimize $D$, the algorithm compromises on a minimal filling ratio
(red dashed line) that does not degrade the reflectivity
significantly. The $h(\beta)$-weighted average reflection at the
optimal filling ratio is around $75\%$.

\begin{figure}[htbp]
  \centering
  \includegraphics[width=0.9\columnwidth]{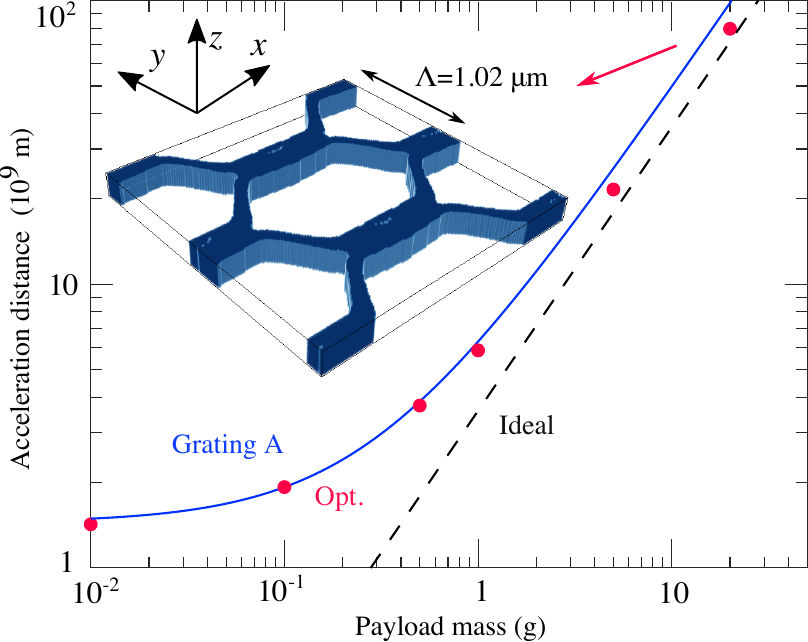}
  \caption{The acceleration distance as a function of payload mass for
    optimized structures (red dots), Grating A of \figref{silicon}
    optimized for a payload mass $0.1$~g (blue solid line), and
    ideal massless perfect reflector (black dashed line). (Inset): the
    optimal structure for payload mass $20$~g.}
  \label{fig:mass}
\end{figure}

Third, at even larger thickness, the emphasis shifts to minimizing the
volume filling ratio for mass reduction, as there is little room for
further improvement of reflectivity that is bounded by $100\%$. Two
strategies are observed in our optimization results: when multiple
layers $N\gtrsim 2$ are to be optimized, the optimal shape is a 1D
grating with reduced thickness obtained by setting several layers to
be vacant; alternatively, when the material is ensured to fill up the
overall thickness by setting $N=1$, the optimal structure switches to
more complicated 2D PhC structures (blue region). For example, the
optimal geometry at thickness $500$~nm and $N=1$, denoted as $B$ and
depicted in the lower \figref{silicon}(b), is a hexagonal lattice that
resembles a trimmed 1D grating, a clear indication of the tendency of
mass reduction. Its reflection spectrum, as shown in
\figref{silicon}(c), reveals the presence of multiple resonance peaks
within the Doppler-shift bandwidth, contributing to the larger average
reflection than that of Grating $A$. However, its FOM is still
outperformed by Grating A.


\begin{figure}[htbp]
  \centering
  \includegraphics[width=1\columnwidth]{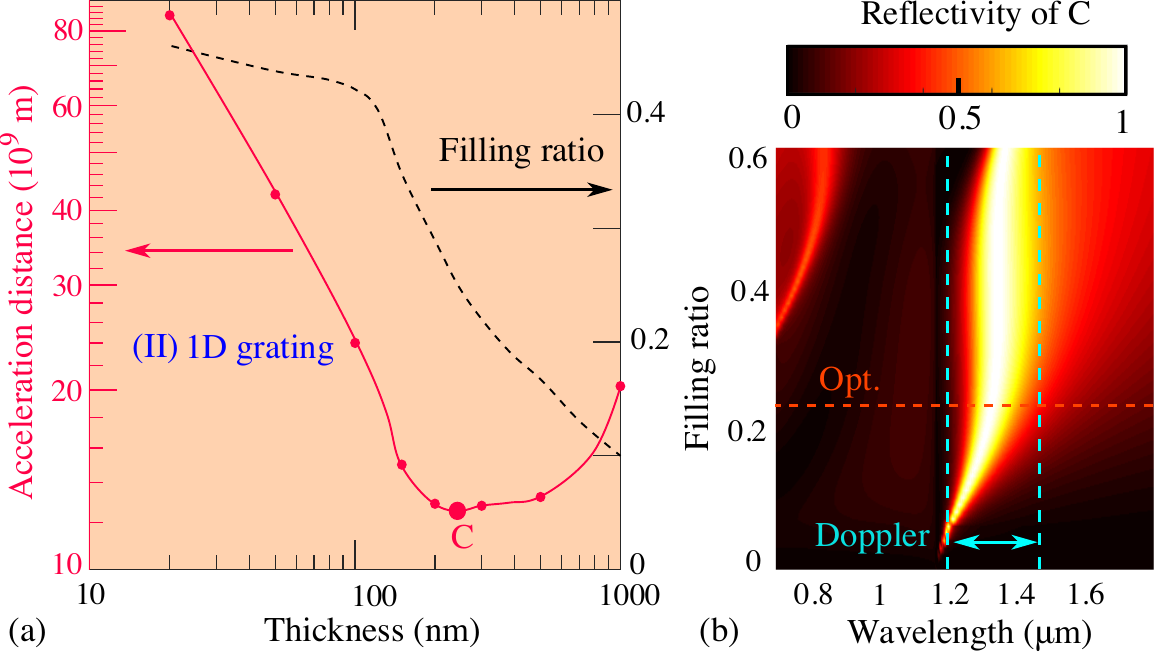}
  \caption{Inverse design of lightsail made of silicon nitride. (a)
    acceleration distance (left axis) and filling ratio (right axis)
    of structures optimized at each thickness, illustrate that over
    the entire range, the optimal geometry is a one-dimensional
    grating (orange). (b) the reflectivity of the optimal grating
    structure $C$ as a function of filling ratio ($w/\Lambda$) and
    wavelength.}
  \label{fig:SiN}
\end{figure}

Next we aim to generalize the above design criteria to other values of
payload mass. At each value of payload mass, we simultaneously
optimize over material distributions, period, and thickness. The FOMs
of the optimal structures (red dots), Grating $A$ (blue solid curve),
and the ideal massless perfect reflector (black dashed line) are
compared in \figref{mass}. Over a wide range of payload mass, the FOM
of Grating $A$ turns out to be very close to the optimal design,
demonstrating that the simple 1D grating is a robust optimal
solution. For a large payload mass, the mass of the sail becomes less
significant and the FOM depends almost entirely on the reflectivity,
thus favoring structures of near-perfect reflection. For instance, for a
payload mass of $20$~g, the optimal structure is a honeycomb lattice
(inset), whose average reflectivity approaches $95\%$.

Finally, we apply these optimization techniques to a lower-index
constituent material, silicon nitride for its appealing mechanical
properties~\cite{moura2018centimeter}. Different structural choice is
expected as lower-index medium reflects light more weakly. As shown in
\figref{SiN} (a), the optimal structure is exclusively a 1D grating
(orange region) throughout the entire range of thickness
$[10,1000]$~nm. The reason for ruling out the two other shapes as
optimal solutions is as follows: at deep subwavelength thicknesses,
reflection increases less dramatically than mass with filling ratio so
that the uniform slabs that obtained for the high index material are
not viable optimized solution here; at larger thicknesses, there is
still much room for improvement of reflectivity, eliminating the need
of cutting down on mass with 2D structuring. The globally optimal
solution, denoted as Grating $C$, occurs at a larger thickness
$243$~nm with a subwavelength period $\Lambda=1.17$~$\mu$m. Its
reflectivity plot as a function of wavelength and filling ratio is
shown in \figref{SiN} (b), exhibiting narrower high-reflectivity band
than the silicon medium. Therefore, similar compromise is made to
decide on a minimal filling ratio, resulting in a moderate average
reflection $57\%$ and larger acceleration distance
$D_{\mathrm{SiN}}\approx13\times 10^9$ m. The degraded FOM might also
be attributed to the large mass density of silicon nitride, which is
around $35\%$ higher than that of silicon. We examine this possibility
by optimizing over a fictitious silicon nitride with the same mass
density of silicon. The optimal structure is found to be almost
identical to that of the real silicon nitride, which leads to slightly
improved $D\approx D_{\mathrm{SiN}}/1.35$, suggesting that the
refractive index plays a more significant role.

\emph{Concluding remarks.} We have developed an optimization framework
that can effectively uncover optimal structure criteria for efficient
lightsail propulsion. Under typical Starshot parameters, the lightsail
geometry obtained by employing our optimization is a 1D subwavelength
grating that outperforms the FOM of previous optimal structures by
almost $50\%$. To ensure that we approach the globally optimal
solutions, we expect future work on deriving a tight theoretical bound
with methods such as Lagrange duality and energy conservation
relations~\cite{molesky2020t}.  We envision that our optimization
framework can be further applied to other challenges in the lightsail
project, including thermal management~\cite{ilic2018nanophotonic} and
propulsion stability~\cite{ilic2019self,siegel2019self}.

\emph{Acknowledgements.} We would like to thank Dr. Zin Lin, Dr. Ian
Williamson, Dr. Momchil Minkov, Dr. Viktar Asadchy, Dr. Bo Zhao, and
Cheng Guo for useful discussions. This work is supported by the
Breakthrough Starshot Initiative, and used the Extreme Science and
Engineering Discovery Environment (XSEDE) supported by the National
Science Foundation, at the San Diego Supercomputing Center through
allocation TG-MCA03S007.

\bibliographystyle{apsrev}
\bibliography{ref}
\end{document}